\documentclass[aps,prl,twocolumn,floats,amsmath,amssymb,showpacs,floatfix]{revtex4}
\usepackage[dvipsnames,rgb,dvips]{xcolor}
\usepackage{epstopdf}

\usepackage{graphicx}
\usepackage{dcolumn}
\usepackage{float}

\usepackage{todonotes}
\usepackage{units}
\newcommand{\ve}[1]{\ensuremath{\mbox{\boldmath$#1$}}}
\newcommand{\ma}[1]{\ensuremath{\mathbb{#1}}}
\newcommand{\ku}{\ensuremath{\mbox{Ku}}}
\newcommand{\sign}{\ensuremath{\mbox{sgn}}}
\newcommand{\st}{\ensuremath{\mbox{St}}}
\newcommand{\tr}{\ensuremath{\mbox{Tr}}}
\newcommand{\rd}{\ensuremath{\mathrm{d}}}
\newcommand\transpose{^{\scriptstyle\mathrm T}}
\newcommand\nn{\nonumber}

\newcommand{\tauK}{\ensuremath{\tau_{\mbox{\tiny K}}}}

\def\clap#1{\hbox to 0pt{\hss#1\hss}}

\newcommand{\eqnlab}[1]{\label{eq:#1}}

\newcommand{\eqnref}[1]{(\ref{eq:#1})}

\newcommand{\Eqnref}[1]{Eq.~(\ref{eq:#1})}
\newcommand{\Figref}[1]{Fig.~\ref{fig:#1}}

\begin{document}
\title{Tumbling of small axisymmetric particles in random and turbulent flows}

\author{K. Gustavsson, J. Einarsson, and B. Mehlig}
\affiliation{Department of Physics, Gothenburg University, 41296 Gothenburg, Sweden}

\begin{abstract}
We analyse the tumbling of small non-spherical, axisymmetric  particles in random and turbulent flows.
We compute the orientational dynamics in terms of a perturbation expansion in the Kubo number,
and obtain the tumbling rate in terms of Lagrangian correlation functions. These capture preferential
sampling of the fluid gradients which in turn can give rise to differences
in the tumbling rates of disks and rods. We show that this is a weak effect in Gaussian random flows.
But in turbulent flows persistent regions of high vorticity cause disks to tumble much faster than
rods, as observed in direct numerical simulations [Parsa {\em et al.}, Phys. Rev. Lett. {\bf 109} (2012) 134501].
For larger particles (at finite Stokes numbers),
rotational and translational inertia affects the tumbling
rate and the angle at which particles collide, due to the formation of rotational caustics.
\end{abstract}

\maketitle

The orientational dynamics of axisymmetric particles in random and turbulent flows is of great significance in many areas of the Natural Sciences and in technology.
For example, turbulent flow visualisation experiments employ reflective flakes \cite{Sch89}.
Patterns of non-spherical particles suspended in flows were investigated in \cite{Wil09,Bez10,Wil11b},
revealing singularities in the orientational patterns of rheoscopic suspensions.
Aerosols in the natural world are often suspensions of small non-spherical particles.
For example, tumbling ice particles in turbulent clouds
may play an important role in cloud-particle interactions \cite{Pru97}.
Small dust grains in circumstellar accretion disks are not spherically symmetric \cite{Pra95,Wil08}.
The relative orientation at which such grains
collide may have important consequences for the outcomes of grain-grain collisions,
and their orientation may have important implications for photophoretic forcing \cite{Eym12}.
A last example concerns plankton in the upper ocean layer.
Their tumbling may influence their nutrient uptake and light scattering \cite{Gua12}.

In all of these cases the particles are smaller than the smallest turbulent eddies in the suspending flow,
and the orientational dynamics of such small particles is driven by the local flow gradients:
the difference in flow velocity over the particle leads to a hydrodynamic torque.
Understanding how small non-spherical particles respond to flow gradients is a necessary step
in describing the collision dynamics of turbulent suspensions of axisymmetric particles.
Morever, the orientational dynamics of small
non-spherical particles is of fundamental interest in turbulence research,
because it reflects the statistics of the velocity gradients in turbulent flows \cite{Pum11}.
Recently, the tumbling rate of small axisymmetric particles in turbulent flows was investigated experimentally and
by direct numerical simulations \cite{Par12}.
It was found that disks tumble, on average, at a much higher rate than rods. This was related to the observation
that rods tend to preferentially align with the vorticity of the flow \cite{Pum11}.
But disks too exhibit alignment with flow structures, the equations of motion for disks and rods are in fact almost the same.  The only difference is that
the flow-gradient matrix for rods is replaced by
its  negative transpose for disks (explained in more detail below).

This raises the questions: which flow configurations are responsible for the 
difference in tumbling between disks and rods?  How do disks align? How does the tumbling in turbulent flows differ from 
that in random flows, how sensitive is the orientational dynamics
to particular features of turbulent flows?
How does the nature of the turbulent Lagrangian flow statistics influence the tumbling?
How does tumbling reflect vorticity?
Finally, what is the effect of particle inertia upon the tumbling?

To answer these questions we analyse the tumbling of small non-spherical particles in random and turbulent flows using perturbation theory.
In the simplest case our problem is governed by three dimensionless parameters.
The Kubo number $\ku=u_0\tau/\eta$ is a dimensionless measure of the correlation time of the flow,
here $u_0$, $\tau$ and $\eta$ are the smallest characteristic speed-, time- and length scales of the flow (Kolmogorov scales in turbulence).
The Stokes number, $\st$,  
characterises the damping of the particle dynamics with respect to the flow.
The third parameter is the aspect ratio $\lambda$ of the axisymmetric particle.

In the limit of $\st\rightarrow 0$, 
the centre-of-mass $\ve r$ is simply advected. The orientational dynamics of the unit vector $\ve n$ pointing
along the symmetry axis of the particle is driven by the local flow gradients
(provided that the dimensions of the particle are much smaller than $\eta$). In other words
$\ve n$ follows Jeffery's equation \cite{Jef22}. We use dimensionless units
 $t=\tau t'$, $\ve r=\eta\ve r'$, $\ve u=u_0\ve u'$. Dropping the primes,
the equation of motion reads:
\begin{align}
\dot{\ve r}&=\ku\,\ve u\,,\quad
\dot{\ve n}=\ku\left[\ma O\ve n+\Lambda\left(\ma S\ve n-(\ve n\transpose\ma S\ve n)\ve n\right)\right]\,.
\eqnlab{eqm_rodn}
\end{align}
Here $\Lambda=(\lambda^2-1)/(\lambda^2+1)$ parameterises the particle shape
($\Lambda = -1$ for disks, $0$ for spheres, and $1$ for rods).
Further $\ma S = (\ma A + \ma A\transpose)/2$ and $\ma O = (\ma A - \ma A\transpose)/2$
are the symmetric and antisymmetric parts of the matrix $\ma A(\ve r_t, t)$ of flow gradients.
The time-averaged tumbling rate  $\langle{{\dot n}^2}\rangle$ is determined by the fluctuations
of $\ma S(\ve r_t,t)$ and $\ma O(\ve r_t,t)$ along the particle trajectories $\ve r_t$.
In the limit of rapidly fluctuating random flows ($\ku \rightarrow 0$) the tumbling rate averaged along trajectories can be replaced by an average over the ensemble of $\ma S$ and $\ma O$.
This average, denoted by $\langle {\dot n}^2\rangle_0$, is determined
by the invariants of  $\ma S$ and $\ma O$. For incompressible, isotropic random flows one finds:
\begin{equation}
\label{eq:rot}
\langle{{\dot n}^2}\rangle \approx \langle {\dot n}^2\rangle_0 =
{\rm Ku}^2 \big(-5\tr\,\langle{\ma O}^2\rangle+
3\Lambda^2\tr\,\langle{\ma S}^2\rangle\big)/15\,.
\end{equation}
Note that in homogenous flows
$\tr\langle\ma S^2\rangle = -\tr\langle\ma O^2\rangle = \tr\langle\ma A\transpose \ma A\rangle/2$.
In turbulent flows $\tr\langle\ma A\transpose \ma A\rangle$  is proportional to the energy dissipation.
An expression equivalent to (\ref{eq:rot}) was first derived in \cite{Shi05} and is also quoted in \cite{Par12}.
Eq.~(\ref{eq:rot}) is symmetric in $\Lambda$, meaning that 
disks tumble at the same rates as rods.
Differences between disks and rods could arise for two reasons. First, one or more symmetries may be broken. For example, breaking isotropy \cite{Vincenzi}
gives rise to an extra term that is odd in $\Lambda$.
Second, 
in homogenous, isotropic, and incompressible flows differences in the behaviour of disks and rods may arise due to preferential sampling of the flow gradients. 

The dynamics of small disks and rods are closely related. Taking the limits $\Lambda\rightarrow -1$ and $\Lambda \rightarrow1$ 
in Eq.~(\ref{eq:eqm_rodn})
shows that the unnormalised orientation vectors $\ve q$  (such that $\ve n = \ve q /|\ve q|$)  of disks and rods obey
$\dot{\ve q}_{\mbox{\tiny disk}}=-\ku\,\ma A\transpose\ve q_{\mbox{\tiny disk}}$
and $\dot{\ve q}_{\mbox{\tiny rod}}=\ku\,\ma A\ve q_{\mbox{\tiny rod}}$.
In persistent flow regions, 
the dynamics of $\ve q$ is determined by
the eigenvectors of $-\ma A\transpose$ or of $\ma A$.
If all eigenvalues of $\ma A$ are real, rods align with the eigenvector corresponding to the largest eigenvalue.
If $\ma A$ has one real and two complex conjugate eigenvalues, rods align if the real eigenvalue is positive and tumble otherwise.
Changing $\ma A\to-\ma A\transpose$ (rods to disks) switches the signs of the eigenvalues. Therefore, when $\ma A$ has complex eigenvalues the transformation $\ma A \to -\ma A\transpose$ alters 
the dynamics from tumbling to aligning, and vice versa. 
The eigenvalues of $\ma A$ can be parameterised by the invariants $\tr \ma A^2$ and $\tr \ma A^3$ \cite{cho90}.
In turbulent flows the joint distribution of $\tr\ma A^2$ and $\tr\ma A^3$ is known to be strongly skewed \cite{Che99}. 
The resulting asymmetry under $\tr \ma A^3\to-\tr\ma A^3$ in the distribution of flow gradients causes different tumbling rates for rods and disks in persistent flows.

This argument explains why rods and disks tumble differently in persistent flow regions, where the flow gradient matrix $\ma A$ remains approximately
constant while $\ve n$ aligns. In turbulent flows, however,  {the matrix $\ma A$ may change on the same time scale as $\ve n$.
The resulting time-dependent problem is very difficult to solve. Below we attack the problem from a different point of view: in the limit of rapidly
changing flows we express the tumbling rate in terms of 
the Lagrangian fluctuations of $\ma A$.}
The orientional dynamics  may be computed by iterating the implicit solution of (\ref{eq:eqm_rodn}):
\begin{align}
\ve n_{t'}&=\ve n_0\!+\!\ku\!\int_0^{t'}\!\!\!\!{\rm d}t\,[\ma O_{t}\ve n_{t}\!%
+\!\Lambda(\ma S_{t}\ve n_{t}\!-\!(\ve n_{t}\transpose\ma S_{t}\ve n_{t})\ve n_{t})]\,.
\eqnlab{n_expansion}
\end{align}
Here $\ma O_t \equiv \ma O(\ve r_t, t)$ and $\ma S_t \equiv \ma S(\ve r_t, t)$. Iteratively substituting
$\ve n_t$ into the r.h.s. of \eqnref{n_expansion}
generates perturbation expansions for $\ve n_t$ and $\dot{\ve n}_t$
in powers of $\ku$.
Averaging 
gives us the leading-order correction to Eq.~(\ref{eq:rot})
\begin{align}
\eqnlab{expansion_general}
&\langle{\dot n^2}\rangle=\langle \dot n^2 \rangle_0
+\frac{2}{5}\ku^3\Lambda\int_0^\infty\!\!\!\!\rd t
\\&
\times\big[ \!-\!\tr\,\langle\ma O_0^2\ma S_{-t}\rangle
\!+\!2\Lambda\tr\,\langle\ma S_0\ma O_0\ma S_{-t}\rangle
\!+\!\frac{3}{7}\Lambda^2\tr\,\langle\ma S_0^2\ma S_{-t}\rangle
\big]\,.\nn
\end{align}
This correction is given
by three-point Lagrangian correlation functions of $\ma O_t$ and $\ma S_t$.

Eq.~(\ref{eq:expansion_general}) shows that the $\ku^3$-correction to Eq.~(\ref{eq:rot}) contains terms antisymmetric in $\Lambda$, causing disks to tumble differently from rods.
For Gaussian random flows, the Lagrangian correlation functions in the integrands of \eqnref{expansion_general}
can be calculated analytically for small $\ku$, as we show below. For turbulent flows we have
determined the correlation functions numerically, using data from the Johns Hopkins Turbulence Database (JHTDB) 
\cite{Yi2008,Yu2012}.

{\em Random flows.}
We represent the incompressible, homogenous, and isotropic random flow
as $\ve u(\ve r,t)=\nabla\wedge\ve A(\ve r,t)$ in terms of a Gaussian random vector potential $\ve A(\ve r,t)$ with zero mean
and correlation function 
$\langle A_i(\ve r_0, 0)A_j(\ve r_0, t) \rangle = \delta_{ij} \exp (-|t|)/6$ \cite{Meh05}.
The corresponding Eulerian correlation functions (evaluated at a fixed point $\ve r_0$ in space) are given by
$\tr\langle\ma S(\ve r_0,0)\ma S(\ve r_0,t)\rangle=-\tr\langle\ma O(\ve r_0,0)\ma O(\ve r_0,t)\rangle=5\,e^{-|t|}/2$.
The Eulerian three-point
functions vanish because the Gaussian gradient distribution is even.     
The Lagrangian correlations at finite values of $\ku$ can be computed perturbatively,
taking into account recursively that the actual trajectory $\ve r_t$  deviates from its initial
condition $\ve r_0$. As shown in \cite{gustavsson2011,Gus13} this yields an expansion in $\ku$.
\begin{figure}[t]
\includegraphics[width=8cm]{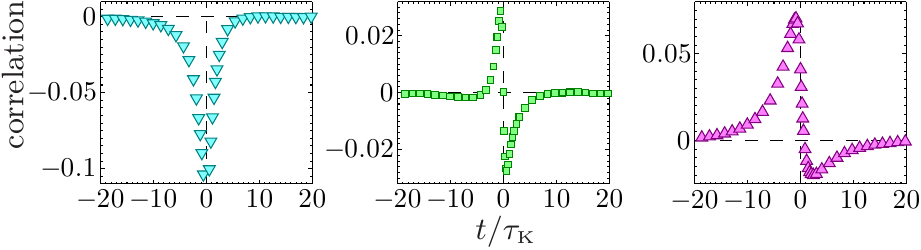}
\caption{\label{fig:1} {\em (Color online)}. Numerical results for Lagrangian correlations $\tr\langle\ma S_{0}^2\ma S_{t}\rangle\tauK^3$ (cyan,$\triangledown$), $\langle\tr\ma S_{0}\ma O_{0}\ma S_{t}\rangle\tauK^3$ (green,$\square$) and $\tr\langle\ma O_{0}^2\ma S_{t}\rangle\tauK^3$ (magenta,$\vartriangle$). Obtained using the JHTDB \cite{Yi2008,Yu2012}, see text.
In units of $\tauK\equiv1/\sqrt{\tr\langle\ma A\transpose \ma A\rangle}$.  }
\end{figure}

The Lagrangian correlation functions quantify the degree of preferential sampling.  
As Eq.~(\ref{eq:expansion_general}) shows differences in tumbling rates between disks and rods are determined by Lagrangian three-point correlations.
We find to third order in $\ku$:
\begin{align}
&\tr\langle\ma S_0\ma O_0\ma S_t\rangle\!=\!\frac{\scriptstyle 35{\rm Ku}^3}{ \scriptstyle 16}\sign(t)e^{-|t|}(1-2|t|e^{-|t|}-e^{-2|t|})\,,\nn\\
&\tr\langle\ma O_0\ma O_0\ma S_t\rangle\! =\! -\frac{\scriptstyle 125{\rm Ku}^3}{\scriptstyle 288}\sign(t)e^{-|t|}(1\!-\!2|t|e^{-|t|}\!-\!e^{-2|t|})\,,\nn
\end{align}
\begin{align}
&\tr\langle\ma S_0^2\ma S_t\rangle\!=\!-\frac{\scriptstyle 175{\rm Ku}^3}{\scriptstyle 96}
\sign(t)e^{-|t|}(1\!-\!2|t|e^{-|t|}\!-\!e^{-2|t|})\,.\label{eq:3point}
\end{align}
Thus the Lagrangian flow-gradient fluctuations are not Gaussian, a consequence of preferential sampling at finite Kubo numbers.
Extending (\ref{eq:expansion_general}) to $\ku^6$ and inserting the required
correlation functions [such as (\ref{eq:3point})] we obtain:
\begin{align}
\label{eq:pt}
&\langle \dot {n}^2\rangle = \frac{\ku^2}{6}(5+3\Lambda^2)-\frac{\ku^4}{4}\Lambda^2(5+3\Lambda^2)\\
&+\!\frac{\ku^6}{864}\Lambda(-25\! +\! 4668\Lambda\!+\!45\Lambda^2\!+\!7236\Lambda^3\!+\!2484\Lambda^5) + \ldots .\nn
\end{align}
Odd powers in $\Lambda$ occur in this expression, giving rise to differences in tumbling between disks and rods.
But the effect is weak, it occurs to order $\ku^6$. {Eq.~(\ref{eq:pt}) (extended to order $\ku^8$) 
can be resummed by Pad\'e{}-Borel
resummation \cite{Gus13a}, yielding accurate results up to $\ku \sim 1$.}

We conclude that disks and rods tumble at almost the same rates in Gaussian random flows. The question is thus what causes
the striking differences between the dynamics of rods and disks observed in \cite{Par12}?

{\em Turbulent flows}. According to Eq.~(\ref{eq:expansion_general}), differences in the tumbling of rods and disks due to preferential
sampling of the flow gradients are parameterised by Lagrangian three-point correlation functions.
We cannot compute these correlation functions analytically, and have thus evaluated them numerically using
the JHTDB \cite{Yi2008,Yu2012}.
The data set contains a direct numerical simulation of forced, isotropic turbulence on a $1024^3$ grid, for circa $45$ Kolmogorov times $\tauK$,
at a Taylor micro-scale Reynolds number $\textrm{Re}_\lambda = 433$.
From the data set we computed the Lagrangian correlations.
The three correlation functions contributing to the tumbling rate in \Eqnref{expansion_general} are shown in
Fig.~\ref{fig:1}. The major contribution after integration comes from the $\tr \langle \ma O_0^2\ma S_{-t}\rangle$-term, with the contribution of the $\tr \langle \ma S_0^2\ma S_{-t}\rangle$-term approximately a factor
$\Lambda^2/3$ smaller. These two terms together result in a substantial contribution to the tumbling rate that is odd in $\Lambda$, giving rise to pronounced differences
in the tumbling of rods and disks.

We have presented arguments valid for persistent flows (large $\ku)$, 
as well as  the perturbative small-$\ku$ result (\ref{eq:expansion_general}). 
We now show that the conclusions drawn from the two arguments
are closely connected.
In turbulence
large positive values of $\tr \ma A^3$ typically coincide with large negative values of $\tr\ma A^2$ (vortex-dominated flow) \cite{Che99,Lut09}.
Conversely, negative values of $\tr \ma A^3$ typically coincide with positive $\tr\ma A^2$ (strain-dominated flow). 
We decompose $\tr\ma A^3=3\tr\ma O^2\ma S+\tr\ma S^3$, and note that in turbulence $\tr\langle\ma O^2\ma S\rangle>0$ and $\tr\langle\ma S^3\rangle<0$ (see Fig.~\ref{fig:1}). 
Thus, large positive values of $\tr \ma A^3$ typically correspond to large values of $\tr\ma O^2\ma S$.
The corresponding flow regions are vortex tubes \cite{She90} that persist long enough for rods to align and disks to tumble.
The differences in the tumbling rates of disks and rods due to such persistent regions are reflected in the integral over $\tr\ma O_0^2\ma S_{-t}$ in \Eqnref{expansion_general}:
{first a local strain creates the vortex tube. This takes several $\tau_{\rm K}$ (Fig.~\ref{fig:1}, right panel).
During this time rods and disks align in different ways, and after that disks start to rotate.}
Strain-dominated flow regions, by contrast, correspond to large negative values of $\tr \ma S^3$ and the differences in tumbling rates result in the integral over $\tr\ma S_0^2\ma S_{-t}$ in \Eqnref{expansion_general}.
In this case the difference between rods and disks is due to the difference in magnitude of eigenvalues. Since the middle eigenvalue of $\ma A$ is positive on average, and the sum of eigenvalues is zero, the magnitude of the first eigenvalue (acting on the rod) is necessarily smaller than that of the last (acting on the disk).  Disks 
respond more quickly than rods to strain-dominated regions and hence exhibit a larger tumbling rate {also in these regions.}

\begin{figure}
\includegraphics[width=8cm]{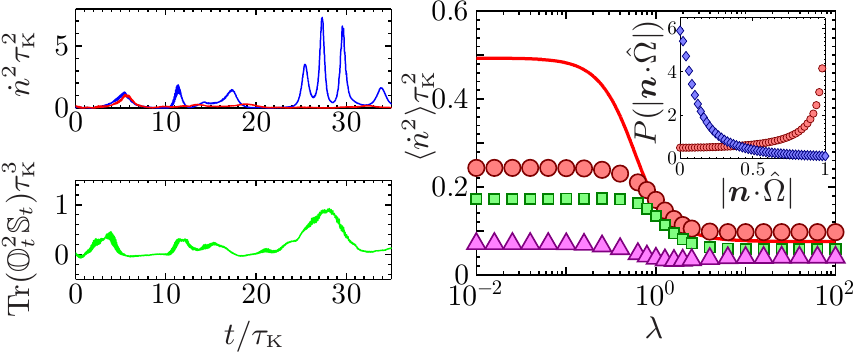}
\caption{\label{fig:2} {\em (Color online)}. Left:
Tumbling rate ${\dot n}^2$ for a disk (blue) and a rod (red) in turbulent flow as a function of time,
using the JHTDB \cite{Yi2008,Yu2012}.
Also shown is
$\tr\,\ma O_t^2 \ma S_t$ (green). Right: average squared tumbling rate $\langle\dot n^2\rangle$ in turbulence
as a function of the aspect ratio $\lambda$ (red,$\circ$).
\Eqnref{expansion_general} with data from \Figref{1} is shown as solid red.
Also shown is ${\dot n}^2$ averaged
conditional on large values of $\tr\,\ma O_t^2\ma S_t$ (green,$\Box$) (23\% of the sampled data) conditional on small values
of $\tr\,\ma O_t^2\ma S_t$ (magenta,$\triangle$) (77\% of the sampled data). Inset: alignment distributions for disks (blue,$\diamond$) and rods (red,$\circ$). }
\end{figure}

Fig.~\ref{fig:2} (left) shows how ${\dot n}^2$ varies as a function of time
in a turbulent flow $\ve u(\ve r,t)$.
Also shown is $\tr \ma O_t^2 \ma S_t$. In agreement with the calculations and arguments
outlined above, rods align and disks tumble strongly when $\tr \ma O_t^2 \ma S_t$ is large and positive.
Fig.~\ref{fig:2} (right) shows numerical results for  $\langle{{\dot n}^2}\rangle$ in a turbulent flow, as a function of 
the aspect ratio $\lambda$.
The numerical simulations are compared to the theoretical result \eqnref{expansion_general} with correlation functions according to \Figref{1}.
\Eqnref{expansion_general} is valid for small $\ku$. In turbulence $\ku\sim 1$, and the numerical tumbling rate $\langle\dot n^2\rangle$ is not expected to  agree quantitatively with \Eqnref{expansion_general}.
{In other contexts we have found that the parameter-dependence (in this case the $\Lambda$-dependence) in $\ku$-expansions is often approximately independent of $\ku$ for small and intermediate $\ku$.  This 
may explain why the general shape of the curve shown in \Figref{2} is approximately correct, but not the amplitude.}

Also shown in \Figref{2} is ${\dot n}^2$ averaged conditional on large $\tr \ma O^2 \ma S$: the substantial difference
in tumbling rates of disks and rods is largely caused by the flow configurations with large $\tr \ma O^2 \ma S$, confirming the picture outlined above.
Finally, when rods align with the leading eigenvector of $\ma A$, then the vorticity vector
$\ve \Omega=({\ve\nabla}\wedge{\ve u})/2$ does the same. 
This is expected since the equations of motion for rods and vorticity have a common term involving $\ma A$ \cite{Wil12,Pum11}.
But for disks $\ve n$ is preferentially orthogonal to $\ve \Omega$ (inset of right panel of Fig.~\ref{fig:2}).

{\em Effects of particle inertia}.
When $\st>0$, different moments of inertia and fluid resistance tensors result in differences in the tumbling
of disks and rods.
Neglecting possible inertial effects due to the fluid, the
dynamics of small spheroidal particles is  given by
\begin{align}
\eqnlab{finiteST_eqm}
\dot{\ve r} &= \ku\,\ve v\,,\qquad \dot{\ve n} = \ku\,{\ve \omega} \wedge \ve n \\\nn
\st\,\dot{\ve v}&=\left[C^{(t)}_\perp\ma I+\left(C^{(t)}_\parallel-C^{(t)}_\perp\right)\ve n\ve n\transpose\right](\ve u-\ve v)\\
\st\,\dot{\ve \omega}&=\left[C^{(r)}_\perp\ma I+\left(C^{(r)}_\parallel-C^{(r)}_\perp\right)\ve n\ve n\transpose\right](\ve\Omega-\ve\omega)
\nn\\&
-\Lambda C^{(r)}_\perp(\ma S\ve n)\wedge\ve n
+\ku\,\st\,\Lambda(\ve n\cdot\ve\omega)\ve\omega\wedge\ve n\nn\,.
\end{align}
Here $\mathrm{St} = m /( 6\pi b \mu \tau)$  is the Stokes number 
of a spherical particle of radius $b$, equal to the minor axis of the spheroidal particle. 
The particle mass is $m$ and $\mu$ denotes the viscosity of the fluid. Further
$\ma I$ is the unit matrix, and the coefficients $C$ are given by translational ($C^{(t)}$) and rotational ($C^{(r)}$) hydrodynamic drag \cite{Bre74} and moment of inertia along ($C_\parallel$) and perpendicular ($C_\perp$) to the particle symmetry axis:
\begin{align}
C^{(t)}_\perp&
=\frac{8(\lambda^2-1)}{3\lambda((2\lambda^2-3)\beta+1)}
\,,\;
C^{(t)}_\parallel
=\frac{4(\lambda^2-1)}{3\lambda((2\lambda^2-1)\beta-1)}
\,,
\nn\\
C^{(r)}_\perp&
=\frac{40(\lambda^2-1)}{9\lambda((2\lambda^2-1)\beta-1)}
\,,\;
C^{(r)}_\parallel
=\frac{20(\lambda^2-1)}{9\lambda(1-\beta)}
\,,\;
\nn\\
\beta&=
\frac{1}{\lambda\sqrt{|\lambda^2-1|}}
\left\{
\begin{array}{ll}
{\rm acos}(\lambda) & \mbox{if }\lambda \le 1 \cr
{\rm acosh}(\lambda) & \mbox{if }\lambda > 1
\end{array}
\right.
\,.
\eqnlab{eqm_coeffs}
\end{align}
Eqs.~(\ref{eq:finiteST_eqm}) and (\ref{eq:eqm_coeffs})
are widely used in the engineering literature, see for example \cite{Mar10} 
and references therein.  In the limit $\st\to0$, \Eqnref{eqm_rodn} is recovered.

\begin{figure}
\includegraphics[width=8cm]{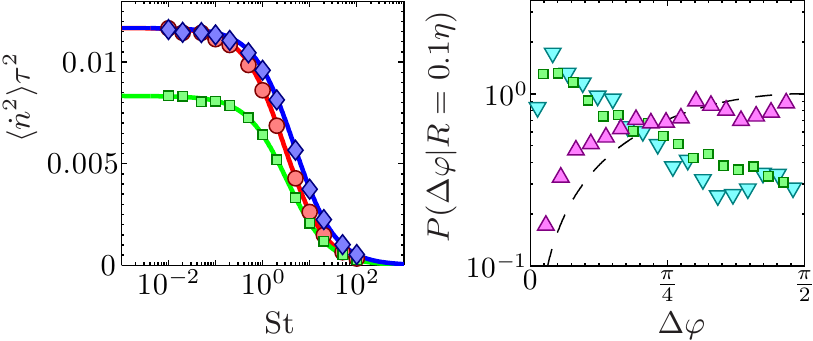}
\caption{\label{fig:3} {\em (Color online)}. Left: $\langle\dot n^2\rangle$ as a function of $\st$ in a Gaussian random flow (see text).
Symbols show results of numerical simulations, solid lines show theory (\ref{eq:stres}). Parameters: $\ku=0.1$, $\lambda=\sqrt{0.1}$ (blue,$\Diamond$), $\lambda=1$ (green,$\Box$), $\lambda=\sqrt{10}$ (red,$\circ$). Right:
distribution of the relative angle $\Delta \varphi$ between the orientation vectors $\ve n$ of two
particles close together (at separation $R=0.1\eta$). Black dashed shows $\sin\Delta\varphi$. Parameters: $\ku=1$, $\lambda=\sqrt{10}$.
$\st=0$ (cyan, $\triangledown$), $\st=1$ (green, $\Box$) and $\st=10$ (magenta, $\vartriangle$).}
\end{figure}

The tumbling rate resulting from \eqnref{finiteST_eqm} can be computed in a small-$\ku$ perturbation theory, analogous to
our treatment of Eq.~(\ref{eq:eqm_rodn}) outlined above. To lowest order
in $\ku$ we find for a spheroid in the random-flow model:
\begin{align}
\langle {\dot n^2}\rangle&\!=\!\frac{\ku^2}{6}\frac{C^{(r)}_\perp\,(5+3\Lambda^2)}{\st+C^{(r)}_\perp}\,.
\label{eq:stres}
\end{align}
The result (\ref{eq:stres}) is shown in Fig. \ref{fig:3} in comparison with results of numerical simulations.
We see that the tumbling rate decreases as $\st$ increases, because the coupling between the flow
and the particle weakens. {Finally,} to order $\ku^2$, translational inertia does not affect the tumbling rate.


{\em Caustics.} At finite Stokes numbers the centre-of-mass motion of inertial particles exhibits caustics \cite{Wil05} where the phase-space manifold
describing the dependence of centre-of-mass velocity upon position folds over. This gives rise to large
velocity differences between close-by particles \cite{Bec10,Gus11,Sal12,Bew13,Fal02}. 
For non-spherical particles phase space contains angular degrees of freedom and caustics cause particles with misaligned orientation vectors to collide. Fig.~\ref{fig:3} (right) shows the distribution of angles $\Delta \varphi$ between orientation vectors of nearby particles.
At larger values of $\st$ caustics occur more frequently, giving rise to a broader distribution of the collision angle $\Delta \varphi$.
At still larger $\st$ the distribution approaches that between uniformly randomly distributed
unit vectors, $P(\Delta\varphi)=\sin\Delta\varphi$.

{\em Conclusions.}
{We have shown in this paper that in the absence of inertial effects,
tumbling in turbulent and random flows is determined, to leading order, by Lagrangian $3$-point correlations of the fluid gradients.
In random flows we have computed these correlations and found that preferential effects exist but  are small.
In turbulent flows we have evaluated the correlation functions numerically, using the JHTDB \cite{Yi2008,Yu2012}. We have found
that they give rise to a substantial difference in the tumbling rate between rods and disks, and have explained this difference
by the fact that persistent regions of high vorticity strongly contribute to the Lagrangian three-point statistics.
For larger particles, we have found that rotational inertia affects the tumbling rate and the angle at which particles collide,
due to the formation of rotational caustics.
It would be interesting to study how the non-ergodic statistics of vortex tubes in turbulent flows affects the tumbling rates of disks and rods with finite but small inertia.}

{\em Acknowledgements}.
Financial support by Vetenskapsr\aa{}det and
by the G\"oran Gustafsson Foundation for Research in Natural Sciences and Medicine are gratefully acknowledged.
The numerical computations were performed using resources
provided by C3SE and SNIC, and the numerical results in Figs.~1 and 2
use data from the Johns Hopkins Turbulence Database.

\end{document}